\documentclass{PoS}
\usepackage{amsmath}
\usepackage{amsfonts,amssymb,latexsym}
\usepackage{amscd}
\usepackage{mathrsfs}
\usepackage{amsthm}
\newtheorem{theorem}{\sffamily\bfseries{Theorem}}[section]

\newtheorem{lemma}[theorem]{\sffamily\bfseries{Lemma}}

\newtheorem{definition}[theorem]{\sffamily\bfseries{Definition}}

\theoremstyle{remark}

\newcommand{\supp}{\mathrm{supp}}

\newcommand{\oR}{{\mathbb R}}

\newcommand{\oC}{{\mathbb C}}

\title{Paley-Wiener-Schwartz Theorem and Microlocal Analysis in Theory of Tempered
Ultrahyperfunctions}

\ShortTitle{PWS Theorem and Microlocal Analysis in Theory of Tempered
Ultrahyperfunctions}

\author{\speaker{Daniel H.T. Franco}\\
        Centro de Estudos de F\'\i sica Te\'orica, Setor de F\'\i sica--Matem\'atica\\
        Rua Rio Grande do Norte 1053/302, Funcion\'arios \\
        Belo Horizonte, Minas Gerais, Brasil, CEP:30130-131.\\
        E-mail: \email{dhtf@terra.com.br}}

\author{Luiz H. Renoldi\\
        Universidade Federal do Esp\'\i rito Santo\\
        Departamento de Matem\'atica\\  
        Campus Universit\'ario de Goiabeiras, Vit\'oria, ES, Brasil, CEP:29060-900.\\
        E-mail: \email{lhrenoldi@yahoo.com.br}}

\abstract{We give some precisions on the Fourier-Laplace transform theorem for tempered
ultrahyperfunctions introduced by Sebasti\~ao e Silva and Hasumi, by considering
the theorem in its simplest form: the equivalence between support
properties of a distribution in a closed convex cone and the holomorphy
of its Fourier-Laplace transform in a suitable tube with conical basis.
We establish a generalization of Paley-Wiener-Schwartz theorem for this setting.
This theorem is interesting in connection with the microlocal analysis, where a
description of the singularity structure of tempered ultrahyperfunctions
in terms of the concept of analytic wave front set is given. We also suggest 
a physical application of the results obtained in the construction and study of
field theories with fundamental length.}

\FullConference{Fifth International Conference on Mathematical Methods in Physics --- IC2006\\
		 April 24-28 2006\\
		 Centro Brasilerio de Pesquisas Fisicas, Rio de Janeiro, Brazil}

\begin{document}

\section{Introduction}
Tempered ultrahyperfunctions were introduced in papers of {\sc Sebasti\~ao e
Silva}~\cite{T} and {\sc Hasumi}~\cite{H}, under the name of tempered
ultradistributions, as the strong dual of the space of test functions
of rapidly decreasing entire functions in any horizontal strip.
Recently, aside from the mathematical interest of the
results presented in Refs.~\cite{T}-\cite{DH}, {\sc Br\"uning--Nagamachi}~\cite{BN}
have conjectured that the properties of tempered
ultrahyperfunctions are well adapted for their use in quantum field
theory with a {\em fundamental length}, while {\sc Bollini--Rocca}~\cite{BR}
have given a general definition of convolution between two arbitrary tempered
ultrahyperfunctions in order to treat the problem of singular pro\-du\-cts
of functions Green also in quantum field theory. The present contribution
contains a brief statement of the results obtained in Ref.~\cite{DH}, where the principal
problem studied was the generalization of Paley-Wiener-Schwartz theorem for the
setting of tempered ultradistributions corresponding to a convex cone and its connection
with the microlocal analysis. 

\section{Tempered Ultrahyperfunctions Corresponding to a Convex Cone}
An open set $C \subset \oR^n$ is called a cone if
$x \in C$ implies $\lambda x \in C$ for all $\lambda > 0$. Moreover,
$C$ is an open connected cone if $C$ is a cone and if $C$ is an open connected set.
In the sequel, it will be sufficient to assume for our purposes
that the open connected cone $C$ in $\oR^n$ is an open convex cone
with vertex at the origin. A cone $C^\prime$ is called compact in $C$ --
we write $C^\prime \Subset C$ -- if the projection ${\sf pr}{\overline C^{\,\prime}}
\overset{\text{def}}{=}{\overline C^{\,\prime}} \cap S^{n-1} \subset
{\sf pr}C\overset{\text{def}}{=}C \cap S^{n-1}$, where $S^{n-1}$ is the unit
sphere in $\oR^n$. Being given a cone $C$ in $x$-space, we associate with $C$ a
closed convex cone $C^*$ in $\xi$-space which is the set $C^*=\bigl\{\xi \in \oR^n
\mid \langle \xi,x \rangle \geq 0, \forall x \in C \bigr\}$.
The cone $C^*$ is called the {\em dual cone} of $C$.

By $T(C)$ we will denote the set $\oR^n+iC \subset \oC^n$. If $C$ is
open and connected, $T(C)$ is called the tubular radial domain in $\oC^n$,
while if $C$ is only open $T(C)$ is referred to as a tubular cone. An
important example of tubular radial domain in quantum field theory
is the {\em forward light-cone}
\[
V_+=\Bigl\{z \in \oC^n \mid {\rm Im}\,z_1 >
\Bigl(\sum_{i=2}^n {\rm Im}^2\,z_i \Bigr)^{\frac{1}{2}}, {\rm Im}\,z_1 > 0 \Bigr\}\,\,.  
\]

We will deal with tubes defined as the set of all points $z \in \oC^n$
such that
\[
T(C)=\Bigl\{x+iy \in \oC^n \mid
x \in \oR^n, y \in C, |y| < \delta \Bigr\}\,\,,
\]
where $\delta > 0$ is an arbitrary number. 

Let $C$ be an open convex cone and let $C^\prime$ be an arbitrary
compact cone of $C$. Let $B[0;r]$ denote a closed ball of the
origin in $\oR^n$ of radius $r$, where $r$ is an arbitrary positive
real number. Denote $T(C^\prime;r)=\oR^n+i\bigl(C^\prime \setminus
\bigl(C^\prime \cap B[0;r]\bigr)\bigr)$. We want to consider the space
consisting of holomorphic functions $f(z)$ such that
\begin{equation}
\bigl|f(z)\bigr|\leq K(C^\prime)(1+|z|)^N e^{h_{C^*}(y)}\,\,,\quad
z=x+iy \in T(C^\prime;r)\,\,,
\tag{{\bf 1}} 
\end{equation}
where $h_{C^*}(y)=\sup_{\xi \in C^*}|\langle \xi,y \rangle|$
is the indicator of $C^*$, $K(C^\prime)$ is a constant that depends
on an arbitrary compact cone $C^\prime$ and $N$ is a non-negative real number.
The set of all functions $f(z)$ which are holomorphic in $T(C^\prime;r)$ and
satisfy the estimate ({\bf 1}) will be denoted by $\boldsymbol{{\mathscr H}^o_c}$.
In what follows, we shall prove two lemmas which will be important for our
extension of Paley-Wiener-Schwartz theorem for the setting of tempered ultrahyperfunctions.

\begin{lemma}
Let $C$ be an open convex cone, and let $C^\prime$ be an arbitrary compact
cone contained in $C$. Let $h(\xi)=e^{k|\xi|}g(\xi)$, $\xi \in \oR^n$, be
a function with support in $C^*$, where $g(\xi)$ is a bounded continuous
function on $\oR^n$. Let $y$ be an arbitrary but fixed point of
$C^\prime \setminus \bigl(C^\prime \cap B[0;r]\bigr)$. Then
$e^{-\langle \xi,y \rangle}h(\xi) \in L^2$, as a function of $\xi \in \oR^n$.
\label{lemma0}
\end{lemma}

\begin{proof}
For details see Ref.~\cite{DH}. 
\end{proof}

\begin{definition}
We denote by $H^\prime_{C^*}(\oR^n;O)$ the subspace of $H^\prime(\oR^n;O)$
of distributions of exponential growth with support in the cone $C^*$:
\begin{equation*}
H^\prime_{C^*}(\oR^n;O)=\Bigl\{V \in H^\prime(\oR^n;O) \mid
\supp(V) \subseteq C^* \Bigr\}\,\,. 
\end{equation*}
\end{definition}

\begin{lemma}
Let $C$ be an open convex cone, and let $C^\prime$ be an arbitrary compact
cone contained in $C$. Let $V=D^\gamma_\xi[e^{h_K(\xi)}g(\xi)]$, where
$g(\xi)$ is a bounded continuous function on $\oR^n$ and $h_K(\xi)=k|\xi|$
for a convex compact set $K=\bigl[-k,k\bigr]^n$. Consider
$V \in H^\prime_{C^*}(\oR^n;O)$. Then $f(z)=(2\pi)^{-n}
(V,e^{-i\langle \xi,z \rangle})$ is an element of
$\boldsymbol{{\mathscr H}^o_c}$.
\label{lemma1}
\end{lemma}

\begin{proof}
For details see Ref.~\cite{DH}. 
\end{proof}

We define ${\mathscr U}_c=\boldsymbol{{\mathscr H}^o_c}/\boldsymbol{\Pi}$ as
being the quotient space of $\boldsymbol{{\mathscr H}^o_c}$ by set of pseudo-polynomials.
Here the set ${\mathscr U}_c$ is the space of tempered ultrahyperfunctions corresponding
to the open convex cone $C \subset \oR^n$. The space ${\mathscr U}_c$ is algebraically
isomorphic to the space of generalized functions ${\mathfrak H}^\prime$. This
result, which represents a generalization of {\sc Hasumi}~\cite[Proposition 5]{H},
was obtained by {\sc Carmichael}~\cite[Theorem 5]{C} in the case where $C$ is an
open cone, but not necessarily connected.

\section{A Generalization of the Paley-Wiener-Schwartz Theorem}
More can be said concerning the functions $f(z) \in \boldsymbol{{\mathscr H}^o_c}$.
It is shown that $f(z) \in \boldsymbol{{\mathscr H}^o_c}$ can be recovered as the
(inverse) Fourier-Laplace transform\footnote{The convention of signs in the Fourier
transform which is used in Ref.~\cite{DH} one leads us to consider the inverse
Fourier-Laplace transform.} of the constructed distribution $V \in H^\prime_{C^*}(\oR^n;O)$.
This result is a generalization of the Paley-Wiener-Schwartz theorem.

\begin{theorem}[PWS-Type Theorem]
Let $f(z) \in \boldsymbol{{\mathscr H}^o_c}$, where $C$ is an open convex cone.
Then the Fourier ultrahyperfunction $V \in H^\prime_{C^*}(\oR^n;O)$ has a uniquely
determined inverse Fourier-Laplace transform $f(z)=(2\pi)^{-n}
(V,e^{-i\langle \xi,z \rangle})$ which is holomorphic in
$T(C^\prime;r)$ and satisfies the estimate
\begin{equation*}
\bigl|f(z)\bigr|\leq K(C^\prime)(1+|z|)^N e^{h_{C^*}(y)}\,\,,\quad
z=x+iy \in T(C^\prime;r)\,\,.
\end{equation*}
\label{PWSTheo} 
\end{theorem}

\begin{proof}
For details see Ref.~\cite{DH}.  
\end{proof} 

\section{Analytic Wave Front Set of Tempered Ultrahyperfunctions}
We shall characterize the spectrum of singularities of tempered ultrahyperfunctions
via the notion of analytic wave front set~\cite{Hor2}. Let us now consider the consequences
of Theorem \ref{PWSTheo}.

\begin{theorem}
If $u \in {\mathscr U}_c(\oR^n)$ and $V \in H^\prime_{C^*}(\oR^n;O)$
$({\mbox{with}}\,\,\,O \subseteq \oR^n)$, then $WF_A(u) \subset \oR^n \times C^*$.
\end{theorem}

\begin{proof}
Let $\{C^*_j\}_{j \in L}$ be a finite covering of closed properly convex
cones of $C^*$. Decompose $V \in H^\prime_{C^*}(\oR^n;O)$ as follows:
\begin{equation}
V=\sum\,V_j\,,\quad{\mbox{such that}}\,\, V_j \in H^\prime_{C_j^*}(\oR^n;O)=
\Bigl\{V_j \in H^\prime(\oR^n;O) \mid \supp(V_j) \subseteq C_j^* \Bigr\}\,\,.
\label{decomp}  
\end{equation}
Next apply the Theorem \ref{PWSTheo} for each $V_j$.
Then the decomposition (\ref{decomp}) will induce a representation of
$u$ in the form of a sum of boundary values of functions $f_j(z)\in
\boldsymbol{{\mathscr H}^o_{c_j}}$, such that $f_j(z) \rightarrow
{\mathscr F}^{-1}[V_j] \in {\mathfrak H}^\prime$ in the strong topology of
${\mathfrak H}^\prime$ as $y={\rm Im}\,z \rightarrow 0$, $y \in C_j^\prime
\subset C_j$. According to Theorem \ref{PWSTheo}, the family of functions
$f_j(z)$ satisfy the estimate
\begin{equation*}
\bigl|f_j(z)\bigr|\leq K(C^\prime)(1+|z|)^N e^{h_{C^*}(y)}\,\,,\quad
z=x+iy \in T(C_j^\prime;r)\,\,.
\end{equation*}
unless $\langle \xi,Y \rangle \geq 0$ for $\xi \in C_j^*$ and $Y \in C_j^\prime$, with
$|Y| < \delta$. Then the cones of ``bad'' directions
responsible for the singularities of these boundary values are contained in
the dual cones of the base cones. So, we have the inclusion
\begin{equation}
WF_A(u) \subset \oR^n \times \bigcup_j\,C_j^*\,\,.
\label{inclusion}  
\end{equation}
Then, by making a refinement of the covering and shrinking it to $C^*$,
we obtain the desired result.
\end{proof} 

\section{Physical Applications of the Results Obtained}
The Paley-Wiener-Schwartz theorem has a conceptual significance for QFT since
it determines the analytic structure of $n$-point correlations functions of the
fields, relating this structure to the support properties implied by the
basic physical notions of causa\-li\-ty and spectral condition. This theorem
underlines the derivation of the main results of the axiomatic QFT, such as
PCT theorem.

In its standard form, the Paley-Wiener-Schwartz theorem deals with the
Fourier-Laplace transform of {\it tempered} distributions and so with
analytic functions which have {\em polynomial growth} at infinity. It enters
in some way or other in the perturbative framework to QFT due to the {\it temperedness}
of the free propagators. However, the
behavior of the fields in a QFT with a fundamental length (as the non-commuta\-ti\-ve
quantum field theories (NCQFT)) can be appreciably {\em more singular}.
This implies that the Wightman framework of local QFT turned out to be too narrow for
theoretical physicists, who are interested in handling situations involving
a QFT with a fundamental length. In particular, for NCQFT some very important evidences
to expect that the traditional Wightman axioms must be somewhat modified are:

\,\,\,

\begin{itemize}

\item {\sl NCQFT are {\bf nonlocal}.}

\,\,\,
\item {\sl The existence of hard infrared singularities in the non-planar sector
of the theory can destroy the {\bf tempered} nature of the Wightman functions}.

\,\,\,
\item {\sl The commutation relations $[x_\mu,x_\nu]=i \theta_{\mu\nu}$
also imply uncertainty relations for space-time coordinates
$\Delta x_\mu \Delta x_\nu \sim \bigl|\theta_{\mu\nu}\bigr|$, indicating that the
notion of space-time point loses its meaning. Space-time points are replaced by cells
of area of size $\bigl|\theta_{\mu\nu}\bigr|$. This suggests the existence of
a finite lower limit to the possible resolution of distance. The {\bf nonlocal}
structure of NCQFT manifests itself in a indeterminacy of the interaction regions,
which spread over a space-time domain whose size is determined by
the existence of a {\bf fundamental length} $\ell$ related to the scale of nonlocality
$\ell \sim \sqrt{\bigl|\theta_{\mu\nu}\bigr|}$}. 

\end{itemize}

\,\,\,Our aim is to apply the results obtained here in the extension of the Wightman
axiomatic approach to NCQFT in terms of tempered ultrahyperfunctions (for details see~\cite{DZH}).
We note that the class of NCQFT in terms of ultrahyperfunctions allows for the possibility
that the off-mass-shell amplitudes can grow at large energies faster than any polynomial
(such behavior {\bf is not possible} if fields are assumed to be tempered only). This fact
is relevant since NCQFT stands as an intermediate framework between string theory and the
usual quantum field theory.


\end{document}